\documentclass[twocolumn]{aastex701}          % for arXiv

\hypersetup{linkcolor=red,filecolor=cyan,urlcolor=magenta}

\newcommand{\laura}[1]{}
\newcommand{\Pranav}[1]{}
\newcommand{\aklant}[1]{}

\usepackage{xcolor}

\newcommand{\msun}{M$_{\odot}$}

\newcommand{\fgas}{$f_{\rm gas}$}
\newcommand{\fgasprog}{$f_{\rm gas,prog}$}
\newcommand{\fgaspm}{$f_{\rm gas,post}$}

\newcommand{\Mstar}{$M_{\star}$}

\newcommand{\illtng}{\texttt{IllustrisTNG}}
\newcommand{\tng}{\texttt{TNG50}}
\newcommand{\tngonehundred}{\texttt{TNG100}}
\newcommand{\tngthreehundred}{\texttt{TNG300}}
\newcommand{\brahma}{\texttt{BRAHMA}}
\newcommand{\sublink}{\texttt{SubLink}}

\newcommand{\lenient}{\texttt{BRAHMA-Lenient}}
\newcommand{\strict}{\texttt{BRAHMA-Strict}}
\newcommand{\lenientrepo}{\texttt{BRAHMA-Lenient-repo}}
\newcommand{\lenientdf}{\texttt{BRAHMA-Lenient-DF}}

\usepackage{amsmath}

\submitjournal{ApJ}

\usepackage{placeins}

\begin{document}

\title{Galaxy merger-driven signatures in massive black hole pair hosts across cosmic time}

% \shorttitle{Merger Signatures in MBH Pair Hosts}
% \shortauthors{Satheesh et al.}

% \title{Host properties of massive black hole pairs across cosmic time}
% \title{Host properties of massive black hole pairs across cosmic time: Enhanced star-formation and black hole accretion}

\author[0000-0003-0963-4686]{Pranav Satheesh}
\email[show]{pranavsatheesh@ufl.edu}
\affiliation{Department of Physics, University of Florida, Gainesville, FL 32611, USA}

\author[0000-0002-7080-2864]{Aklant K. Bhowmick}
\email[show]{whm4dg@virginia.edu}
\affiliation{Department of Astronomy, University of Virginia, 530 McCormick Road, Charlottesville, VA 22904}
\affiliation{Virginia Institute for Theoretical Astronomy, University of Virginia, Charlottesville, VA 22904, USA}
\affiliation{The NSF-Simons AI Institute for Cosmic Origins, USA}
\author[0000-0002-2183-1087]{Laura Blecha}
\email{lblecha@ufl.edu}
\affiliation{Department of Physics, University of Florida, Gainesville, FL 32611, USA}

\begin{abstract} 
Galaxy mergers can drive gas toward galactic nuclei, enhancing both star formation and black hole accretion in post-merger galaxies. Here we use the \texttt{IllustrisTNG-50} simulation and two boxes from the \brahma\ simulation suite, all of which adopt different black hole formation models, to characterize merger-associated signatures in candidate massive black hole (MBH)-pair hosts. In all three simulations, we select post-merger galaxies whose progenitors both host MBHs and quantify their enhancements in star formation and black hole accretion relative to matched non-merging controls. In \tng, we find substantial merger-triggered enhancements in both specific star formation rate (sSFR) and specific black hole accretion rate (sBHAR) at $0 \leq z < 3.3$, reaching $\sim1.4 \times$ and $\sim1.5-1.7 \times$ the controls, respectively. The \brahma\ runs employ more physically motivated seed formation models, seeding BHs in lower-mass galaxies and at higher redshifts than \tng. In \brahma, merger-driven enhancements in sSFR and gas fractions are comparable to those in \tng\ at $z \lesssim 3.3$ and persist to the highest redshifts probed by \brahma\ ($z\sim6$). Despite this, \brahma\ shows no significant merger-driven enhancement in sBHAR across its entire redshift range. In a similar vein, \tng\ pair hosts show an enhanced AGN fraction relative to matched controls, whereas \brahma shows only a weak excess among the most rapidly accreting MBHs.
\end{abstract}


\section{Introduction}
\label{sec:intro}
Mergers are a fundamental piece of galaxy evolution and hierarchical structure formation in the $\Lambda \rm CDM$ paradigm. Theoretical studies have shown that tidal interactions from galaxy mergers can lead to bar instabilities, and strong gravitational torques drive cold gas into the galaxy center, triggering starbursts and enhancing active galactic nuclei (AGN) activity \citep{Barnes_and_Hernquist_1991,Mihos_and_Hernquist_1994,Mihos_and_Hernquist_1996,DiMatteo2005,Springel2005a,Springel2005b,Springel_and_Hernquist_2005,Capelo2015,Capelo2017, Moreno2015,Moreno2019}. Observational studies generally support the predictions of isolated galaxy merger simulations, especially at low redshift. Local observations of galaxy close pairs and post-merger galaxies in the Sloan Digital Sky Survey (SDSS) have shown enhanced star formation rates (SFR) \citep{Ellison2008, Scudder2012, Patton2013, Ellison2013, Thorp2019}, morphological disturbances \citep{Casteels2013,
Patton2016}, and enhanced molecular and atomic gas reservoirs \citep{Ellison2018, Violino2018}. However, the merger--AGN connection is more contested. Some studies find an excess AGN fraction in interacting systems \citep{Ellison2011,Satyapal2014,Weston2017, Ellison2019} while others report no enhancement relative to non-interacting control galaxies \citep{Cisternas2011,Kocevski2012,Villforth2014,Mechtley2016,Marian2019}. 

The enhancement of star formation and AGN activity driven by galaxy mergers has been studied extensively across several simulation approaches. Isolated galaxy merger simulations have shown that star formation rate peaks near the first pericentric passage \citep{Hopkins2008,Renaud2014} with black hole accretion rate peaking closer to the final coalescence \citep{Springel2005a,Hopkins2012}. Higher resolution simulations of merging pairs \citep{Moreno2019} have found  SFR elevated by $\sim$30\% relative to isolated counterparts. In the zoom-in simulations of \citet{Dubois2015}, a major gas-rich merger triggers black hole (BH) growth in runs with efficient supernova feedback. \citet{Sharma2024} find that AGN activity is not primarily merger-driven except in the low-redshift regime, and \citet{Jhee2026} reach similar conclusions by analyzing JWST mock observations of merging galaxies from a zoom-in simulation. Disagreement between these simulation studies reflects not only differences in resolution but also the variety of subgrid models used. 

While isolated and zoom-in simulations achieve higher resolution and capture a realistic interstellar medium, they lack the volume necessary to provide statistically meaningful samples to study merger-based enhancements. Cosmological simulations fill this gap by allowing us to probe the merger--SFR and merger--AGN connections across cosmic time. Enhanced specific star formation rates (sSFR, the star formation rate per unit stellar mass) are seen in both interacting pairs and post-merger galaxies in \illtng\ \citep{Nelson2019, Pillepich2018}, with \citet{Patton2020} finding a mean enhancement of a factor of $\sim2$ at the smallest separations and \citet{Hani2020} measuring a comparable sSFR enhancement in post-mergers that decays over $\sim 0.5$~Gyr following coalescence. Enhanced AGN activity and specific black hole accretion rates (sBHAR, black hole accretion rate per unit black hole mass) are likewise found in interacting galaxies relative to isolated ones in \illtng\ \citep{Bhowmick2020,Byrne_Mamahit_2023,Byrne_Mamahit_2024}. Using \tng, \citet{Schechter2025} report excess sSFR in merger hosts out to $z\sim3$ and enhanced sBHAR out to $z\lesssim2$, with the
enhancement weakening toward higher redshifts.

SFR enhancements are also observed in SIMBA mergers at $0<z<2.5$ \citep{Montero2019}. However, at similar redshift ranges in the Horizon-AGN simulation  \citet{kaviraj_galaxy_2015} find that major mergers elevate star formation by only $\sim20$--$40$\% and are not the dominant driver of stellar mass growth across the population. Further, \citet{Martin2017} showed that star formation directly attributable to mergers is small at all redshifts, with mergers being more effective at star formation enhancement at $z<1$. In \textsc{EAGLE} \citep{Schaye2015, Crain2015}, \citet{McAlpine2020} find an over-abundance of luminous AGN in disturbed systems with $M_\star \sim 10^{10}$~\msun\ at $z<1$. \citet{Steinborn2018} using Magneticum Pathfinder simulations find merging systems make up less than 20\% of AGN hosts, hence mergers play a minor role in driving the overall AGN
population. 

The results from simulations paint a picture in which the merger--SFR and merger--AGN connections are strongest at low redshift. Galaxies generally have higher gas fractions, higher star formation rates and more turbulent disks at high-$z$ \citep{Madau2014,Ubler2019}. At these redshifts, secular processes may be more important than mergers in fueling AGN activity. Such processes include bar instabilities \citep{Kormendy_and_Kennicutt_2004,Hirschmann_2012}, clumpy unstable gas-rich disks \citep{Bournaud2011}, cooling flows from the hot halo \citep{Nulsen2000}, and recycled gas from stellar evolution in galactic nuclei \citep{Ciotti_and_Ostriker_1997, Ciotti_and_Ostriker_2007}. 

The advent of JWST has made high-redshift mergers easier to probe. \citet{Puskas2025} analyze close pairs in JADES and find a significant sSFR excess at the smallest separations, strengthening toward $z=5$--$9$, but no significant AGN excess over $z=3$--$9$. However, a close-pair selection does not probe the post-coalescence phase in which AGN triggering may be strongest. \citet{Duan2026} reach a similar conclusion across eight JWST observational fields at $z=4.5$--$8.5$, reporting merger-enhanced star formation but a weak AGN excess. Using JWST/NIRCam imaging of spectroscopically confirmed galaxies at $5.0<z<5.6$, \citet{Omori2026} found that interaction-induced starbursts contribute significantly to stellar mass assembly at this epoch.

In this work we probe the merger--SFR and merger--AGN connections in post-merger galaxies selected as candidate MBH-pair hosts from $z=0$--$6$ using \tng\ and the \brahma\ simulation suite \citep{Bhowmick2024a, Bhowmick2024b, Bhowmick2025a, Bhowmick2025b, Bhowmick2026}. \brahma\ implements physically motivated BH seed models, allowing us to test the sensitivity of merger-driven signatures to seeding physics at high redshift. As \brahma\ seeds black holes in halos some three orders of magnitude less massive than the \illtng\ threshold, it attains a far higher BH occupation fraction at high redshifts and therefore yields many more MBH pairs available to merge at early times \citep{Bhowmick2024a, Bhowmick2026}. The combination of our cosmological simulation suite therefore lets us probe the characteristics of MBH pair hosts across cosmic time. 

The MBH pair hosts studied here are potential precursors of sources for low-frequency gravitational-wave (GW) detectors, particularly the Laser Interferometer Space Antenna \citep[LISA;][]{Amaro_Seoane_2023} and pulsar timing arrays (PTAs), which have reported evidence for a nanohertz stochastic GW background consistent with a cosmic population of massive black hole binaries \citep{Agazie2023a, Agazie2023b}. Merger-enhanced star formation and AGN activity are directly imprinted on the host galaxy shortly after a galaxy merger. Connecting these signatures to the hosts of GW sources detectable by PTAs and LISA additionally requires modeling the subsequent MBH inspiral and coalescence. More broadly, this study helps us understand how galaxy mergers fit into the picture of BH--galaxy coevolution.

The paper is outlined as follows. In Section~\ref{sec:data} we go over the simulation data used in this work, followed by Section \ref{sec:methods} where we describe how we select the MBH pair host samples and match them to non-merging controls. The results are outlined in Section~\ref{sec:results}, which covers the sSFR and sBHAR enhancement, AGN excess and the impact of these merger enhancements on the gas fraction, stellar mass, BH mass and mass ratio. In Section~\ref{sec:discussion} we discuss and compare our results to the literature on merger enhancements in simulations and in observations. We finally summarize our findings and discuss caveats and future work in Section~\ref{sec:summary}.

\section{Data}
\label{sec:data}

\subsection{IllustrisTNG}
\label{sec:illtng}
\illtng\ is a set of cosmological simulation runs performed with the moving-mesh gravity $+$ magnetohydrodynamic (MHD) code \textsc{Arepo} \citep{Springel2010, Pakmor2011,Pakmor2016,Weinberger2020} in three cubic volumes with side 
lengths of roughly 50, 100, and 300 Mpc, referred to as \tng, \tngonehundred, and \tngthreehundred, respectively \citep{Springel2018,Pillepich2018,Nelson2018,Marinacci2018,Naiman2018,Nelson2019,Pillepich2019}. This work utilizes \tng, the highest-resolution realization of the $[51.7\, {\rm Mpc}]^3$ box sampled by $2\times2160^3$ gas and dark matter resolution elements, corresponding to a baryonic mass resolution of $8.5 \times 10^{4}$ \msun\ and a dark matter (DM) resolution of $4.5 \times 10^5$ \msun. \illtng\ adopts cosmological parameters from \citet{Planck2016} and follows the self-consistent evolution of DM, gas, stars, and supermassive black holes from $z=127$ to $z=0$. Halos are identified using the Friends-of-Friends (FOF) algorithm \citep{Davis1985} using a linking length of 0.2 times the mean inter-particle separation, and subhalos are found with \textsc{SUBFIND} \citep{Springel_2001} at each snapshot. The gravitational softening length for the collisionless components (DM and stars) is $288$ pc at $z=0$, and star-forming gas cells have average sizes of $70-140$ pc. The spatial resolution of \tng\ thus allows us to resolve the internal structure of the star-forming disks, including disk instabilities compared to other \illtng\ volumes.

\tng\ includes gas radiative processes, stellar evolution and chemical enrichment, black hole formation, growth and feedback. Star formation follows \citet{Springel_and_Hernquist_2003} and \citet{Vogelsberger2013}. Gas cells above a density threshold of $\rho_{\rm sfr} = 0.13 \, \rm cm^{-3}$ are stochastically converted into stellar particles on a timescale of $t_{\rm sfr} = 2.2 \, \rm Gyr$. Each stellar particle represents a single stellar population described by a \citet{Chabrier2003} initial mass function and evolves according to \citet{Vogelsberger2013}, with modifications described in \citet{Pillepich2018}. Stellar populations redistribute metals back into the interstellar medium (ISM) via stellar and Type Ia/II supernova feedback, modeled as galactic-scale winds \citep{Pillepich2018b}.

BHs are seeded with an initial mass of $1.2 \times 10^{6}$ \msun\ in any FOF halo that does not already host a black hole and that exceeds a total mass of $M_{\rm FOF} = 7.4 \times 10^{10}$ \msun. BHs grow via accretion and mergers. Accretion proceeds at the Eddington-limited Bondi--Hoyle rate, where the Bondi accretion rate is

\begin{equation}
    \dot{M}_{\rm Bondi} = \frac{4 \pi G^2 M^2_{\rm BH} \rho}{c^3_s}
    \label{eq:Bondi}
\end{equation}

\noindent the Eddington rate is

\begin{equation}
    \dot{M}_{\rm Edd} = \frac{4 \pi G M_{\rm BH} m_p}{\epsilon_r\sigma_T c} 
    \label{eq:Eddington}
\end{equation}

\noindent and the black hole accretion rate (BHAR) is therefore

\begin{equation}
    \dot{M}_{\rm BH} = \min\!\left(\dot{M}_{\rm Bondi},\,\dot{M}_{\rm Edd}\right),
    \label{eq:BHAR}
\end{equation}

\noindent where $\rho$ and $c_s$ are the gas density and sound speed in the vicinity of the black hole, $\epsilon_r=0.2$ is the radiative accretion efficiency, and $\sigma_T$ is the Thomson cross-section. The Eddington ratio $\lambda_{\rm Edd}$ is defined as the ratio of $\dot{M}_{\rm BH}$ to $\dot{M}_{\rm Edd}$.

\tng\ employs two BH feedback modes: a thermal mode and a kinetic mode \citep{Weinberger2017}. When the Eddington ratio exceeds a threshold $\chi = \min[0.002\,(M_{\rm BH}/10^8\, M_{\odot})^2,\,0.1]$, the \emph{thermal} (high-accretion) mode is active. The energy is injected continuously and uniformly into the surrounding gas which can drive outflows. Below this threshold, the \emph{kinetic} (low-accretion) mode operates and energy is injected stochastically as time-pulsed winds with a randomly reoriented direction at each event \citep{Nelson2019, Weinberger2017}. As BH dynamics are not resolved in \tng, a repositioning scheme pins each black hole to the local potential minimum, preventing artificially large kicks from neighboring DM particles. As a result, BHs are merged promptly without any merger delay.

\subsection{BRAHMA}
\label{sec:brahma}

The \brahma\ simulation suite \citep{Bhowmick2024a,Bhowmick2024b,Bhowmick2025a,Bhowmick2025b,Bhowmick2026} uses the same \textsc{Arepo} code, the same cosmology and the same halo and subhalo identification as \illtng\ described in Section \ref{sec:illtng}. The initial conditions were generated with the \textsc{MUSIC} algorithm \citep{Music2011} at $z=127$. The \brahma\ runs used here are $[36.9 \, \rm Mpc]^3$ boxes with $1024^3$ DM particles and an equal number of gas cells, with DM and gas mass resolutions of $1.5 \times 10^6$ \msun\ and $2.4 \times 10^5$ \msun\ and a comoving gravitational softening length of $0.73$ ckpc. At this mass resolution, these boxes cover a larger volume than the previously published \brahma\ simulation runs, and have been run to $z=1.3$. Apart from the  BH seeding and dynamics described below, \brahma\ simulations use the full \illtng\ galaxy formation model. 

\subsubsection{BH subgrid models}

The key feature of \brahma\ is its physically motivated BH seeding prescription. These runs use a minimum BH seed mass of $M_{\rm seed} = 1.5 \times 10^5$ \msun\ and combine the following seeding criteria, each motivated by direct collapse black hole (DCBH) \citep{bromm2003,begelman2006,Lodato2006,volonteriEvolutionMassiveBlack2007} formation channels: 

\begin{itemize}
    \item \emph{Halo mass criterion}: BH seeds are placed only within resolved subhalos, requiring a minimum of 32 DM particles.
    \item \emph{Dense and metal-poor gas mass criterion}: The host halo must contain a minimum gas mass of $5 \times M_{\rm seed}$ in dense, star-forming ($\rho \geq 0.13 \, \rm cm^{-3}$) and metal-poor ($\leq 10^{-4} Z_{\odot}$) gas.
    \item \emph{Lyman--Werner flux criterion}: The dense, metal-poor gas must be exposed to a minimum amount of Lyman--Werner (LW) flux of $J_{\rm crit} = 10\,J_{21}$, computed using the analytical formalism from \citet{Dijkstra2014}.
    \item \emph{Gas-spin criterion}: The host halo must have a spin parameter below the Toomre instability threshold, following \citet{Lodato2006}.
    \item \emph{Halo environment criterion}: Seeds are placed only in halos residing in rich environments, defined as having at least one neighbor of comparable mass within a distance of 5 times its virial radius.
\end{itemize}

Each of our \brahma\ boxes applies a different combination of the above seeding criteria. The most lenient seed model run applies the \emph{halo mass criterion}, the \emph{dense and metal-poor gas criterion} and the \emph{LW flux criterion}, and will be referred to as \lenient. We also run another box with a restrictive seed model that applies the additional \emph{gas-spin criterion} and \emph{halo environment criterion}. This run will be referred to as \strict. Additionally, for both boxes we include a subgrid dynamical friction (DF) model. This subgrid model accounts for the dynamical friction force that is unresolved at our mass resolution, using the \citet{Ma2021} correction to the black hole acceleration. Because dark matter particles are $\sim10$ times more massive than gas particles and can artificially heat low-mass black holes, seeds in the DF runs are initialized with an enhanced \emph{dynamical mass} of $24\,M_{\rm seed}$. The dynamical mass ($24\,M_{\rm seed}$) is used only in the gravity calculation, while the actual BH seed mass ($M_{\rm seed}$) is unchanged for BH accretion and feedback. The BHs are merged when their separation falls below twice the gravitational softening length (1.46 ckpc) and they are gravitationally bound, i.e., $|\Delta \mathbf{v}|^2/2 + \Delta \mathbf{a} \cdot \Delta \mathbf{r} < 0$, for the relative velocity ($\Delta \mathbf{v}$), acceleration ($\Delta \mathbf{a}$) and displacement ($\Delta \mathbf{r}$) of the BH pair \citep{Bhowmick2026}.

\section{Methods}
\label{sec:methods}
\subsection{Selecting subhalos hosting MBHs}
\label{sec:subhalo-selection}

We identify merging and non-merging subhalos in \tng\ and \brahma\ using \sublink\ merger trees \citep{RodriguezGomez2015}. A subhalo merger is defined for any case in which a subhalo has more than one direct progenitor, and we refer to the descendant of the two merging subhalos as the post-merger galaxy. We restrict our analysis to mergers in which both progenitors host at least one BH, and select their post-merger descendants as candidate MBH-pair hosts. Throughout this work, we use ``MBH pair hosts" or simply ``pair hosts" to refer to this population. Descendants may contain multiple MBH particles or a numerically merged BH remnant but this selection does not require a resolved or gravitationally bound MBH pair at the descendant's snapshot. Subhalos that do not have more than one direct progenitor in the preceding snapshot are classified as non-mergers, and form the pool from which we draw controls. We further require these non-mergers to host at least one MBH, so that BH-related properties such as the accretion rate can be compared directly against the pair hosts.

To ensure that the MBH pair hosts are well resolved for measuring star formation and black hole accretion rates, we require \tng\ MBH pair hosts and their progenitors to contain at least 1000 stellar particles. We have verified that the results are insensitive to a looser stellar-particle threshold. The requirement that both progenitors host a BH is far more restrictive for \tng, where of the $15{,}383$ galaxy mergers in which both progenitors are above this threshold, only $504$ ($3.3 \%$) have a BH in both progenitors. This cut acts almost entirely on the less massive progenitor and is a consequence of \tng\ seeding BHs only in halos above $7.4\times10^{10}\,\rm M_\odot$. Since the \brahma\ simulations allow black hole seeds to form in much smaller halos than the \tng\ seeding threshold, many of the BH hosts are not well resolved despite the comparable mass resolution of the two suites. We therefore adopt a looser cut of at least $100$ stellar particles for \lenient\ and \strict. The BH requirement is also restrictive in these runs, with $52.9\%$ of mergers retained in \lenient\ and $33.3 \%$ in \strict. For the non-merging galaxies from which controls are drawn, it retains $36.4 \%$ of galaxies of the \tng\ non-merger pool, $91.6\%$ in \lenient\ and $35.5\%$ in \strict.

We further exclude post-merger galaxies that are currently interacting with another nearby companion, in order to isolate the effect of the recent merger from that of an ongoing interaction. Following \cite{Patton2020} and \cite{Hani2020}, we define the dimensionless pair separation

\begin{equation}
    r_{\rm sep} = \frac{r}{R^{\rm host}_{1/2} +R^{\rm comp}_{1/2} }
\end{equation}

\noindent where $r$ is the separation between the host galaxy and its nearest neighbor companion. $R^{\rm host}_{1/2}$ is the stellar half-mass radius of the post-merger galaxy and $R^{\rm comp}_{1/2}$ is the stellar half-mass radius of its nearest neighbor. We discard any post-merger galaxy with a companion at $r_{\rm sep} \leq 2$. The same requirement is applied to the non-merging control pool before matching.

An important caveat is that neither \tng\ nor \brahma\ follows the MBH inspiral to coalescence. In \tng, repositioning pins the MBHs to the local potential minimum and MBHs are promptly merged once the merger criterion is satisfied. In \lenient\ and \strict, the MBHs sink under the subgrid DF force. In both cases the sub-kpc MBH inspiral is unresolved, so the post-merger galaxies may contain a numerically merged BH remnant. Such a numerically merged remnant may correspond to a physical MBH pair that has not yet coalesced. The merger timescale requires additional modeling of the MBH binary hardening \citep{kelley_massive_2017,sayeb_massive_2021,satheesh2025}. We therefore interpret the selected post-merger galaxies as candidate MBH-pair hosts and will explore the effect of MBH merger timescales in future work. Additionally, the subgrid DF model in the \brahma\ runs produces a population of wandering BHs, also seen in other simulations \citep{ricarte_unveiling_2021,Weller2023}. Therefore when describing BH-specific properties in this work, such as sBHAR, we use the most massive BH in the subhalo. We have verified that using the most rapidly accreting BH instead does not change our main conclusions regarding sBHAR.

The cadence of the simulation snapshots is also an important factor in the selection process. Merger-triggered enhancements are stronger closer to coalescence and weaken afterwards \citep{Hani2020,Schechter2025}. For example, \citet{Hani2020} find that post-merger sSFR enhancement in \illtng\ boxes decays over $\sim 0.5$\,Gyr. Since we probe our MBH pair hosts at the snapshot immediately following a subhalo merger, the time interval between successive snapshots ($\Delta \rm T$) sets how far past the merger our measurement of the host property is made. All of the \tng\ snapshots are spaced at $\Delta \rm T < 0.25$\,Gyr, and the \brahma\ runs used here have a finer cadence than earlier \brahma\ runs \citep[e.g.,][]{Bhowmick2026}, with $\Delta \rm T$ remaining below $0.35$\, Gyr. As the snapshot spacing in our simulations is always shorter than $0.5$\,Gyr, every MBH pair host has its properties measured well inside the window over which merger-driven enhancements are expected to persist. This timescale also raises the possibility that control galaxies with mergers more than one snapshot earlier retain some of the merger enhancements. We have checked that excluding control galaxies that experienced a merger within the preceding $0.5$ Gyr only causes minor changes in the measured properties and no change to our main conclusion.

\subsection{Finding control matches to MBH pair hosts}

\begin{figure}[!htb]
    \centering
    \includegraphics[width=0.9\linewidth]{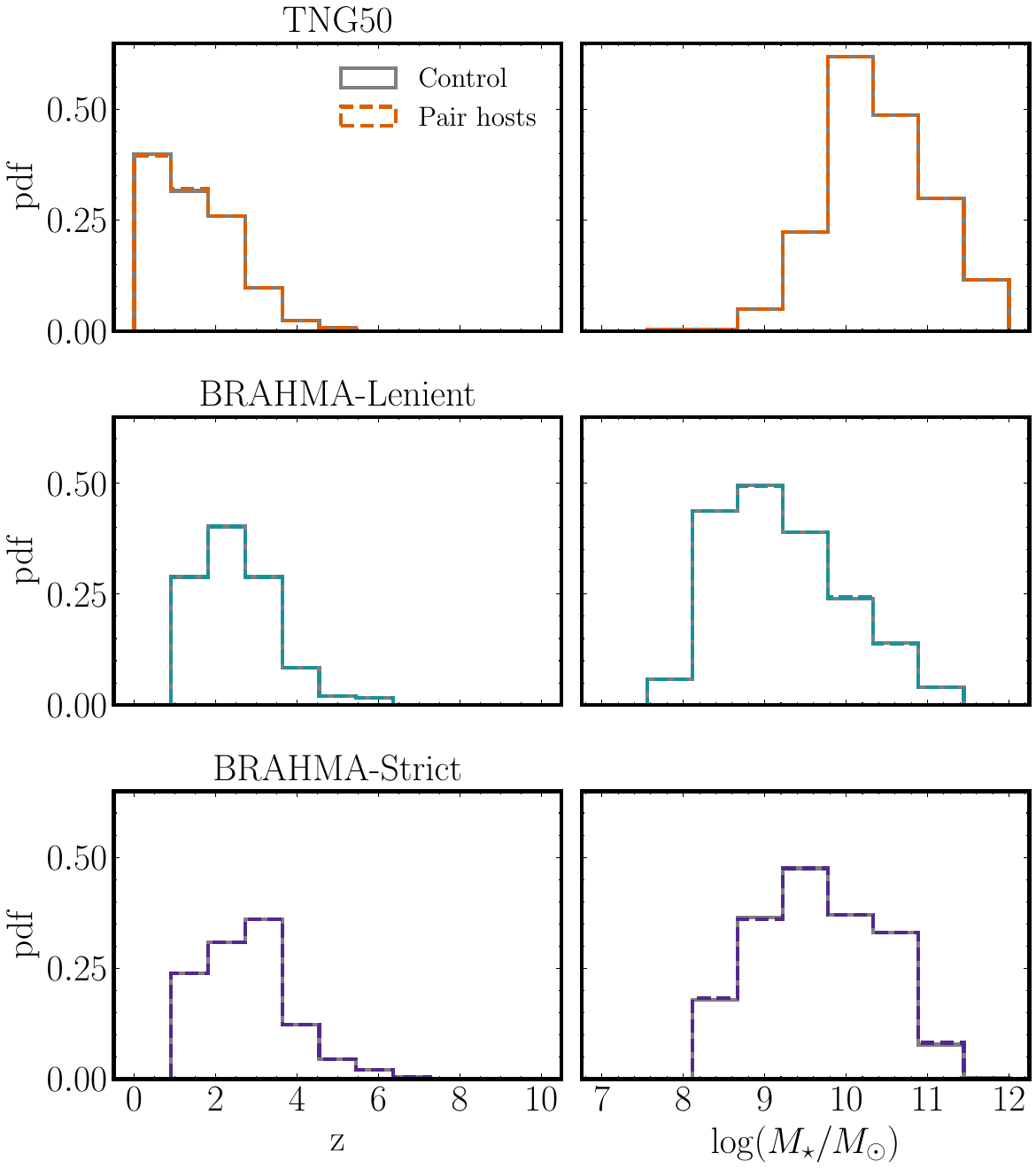}
    \caption{The redshift and stellar mass distributions of MBH pair hosts and their controls for all the simulations used in this work. The Hellinger distances corresponding to the matches are reported in Table \ref{table:sim-details}. The matched controls closely reproduce the redshift and stellar-mass distributions of the selected post-merger hosts in all three simulations.}
    \label{fig:Mstar-z-match-plot}
\end{figure}

We construct a control sample population by matching each pair host to a non-merging galaxy in redshift ($z$) and stellar mass (\Mstar). For all three runs (\tng, \lenient\ and \strict) we select the closest matching non-merger within tolerances of $\Delta z = 0.1$ and $\Delta \log M_\star = 0.1$\,dex. Once a non-merger is assigned as the control for a given pair host, it is removed from the pool of available controls for other pair hosts. This ensures a unique one-to-one match for every pair host. This matching procedure follows a similar approach used in \cite{Schechter2025}. The $z$ and \Mstar\ distributions of the MBH pair hosts and their controls are shown in  Figure~\ref{fig:Mstar-z-match-plot}.

To quantify the quality of the matches, we compute the Hellinger distance between the pair-host and control distributions of redshift and stellar mass. For two discrete distributions $P=\{p_i\}$ and $Q=\{q_i\}$, defined on a common set of $k$ bins, the Hellinger distance is defined as

\begin{equation}
    H(P,Q) = \frac{1}{\sqrt{2}} \sqrt{\sum_{i=1}^{k} (\sqrt{p_i} - \sqrt{q_i})^2}.
    \label{eq:Hellinger}
\end{equation}

\noindent which takes values in $[0,1]$. $H=0$ indicates identical distributions and $H=1$ indicates disjoint distributions. For \tng, the redshift Hellinger distance is $0.061$, while the \brahma\ runs have a matching control at the same snapshot as each pair host, making the Hellinger distances zero. The stellar mass matches are likewise good, with $H(M_\star) \lesssim 0.06$ for all three simulations. Table~\ref{table:sim-details} lists the Hellinger distances for the redshift and stellar-mass matches, along with the number of pair hosts with a matched control. We also classify the pair hosts by the stellar mass ratio ($\mu_\star$) corresponding to their progenitor galaxy mergers. Major mergers are defined as $\mu_\star \geq 0.25$, while $0.1 \leq \mu_\star < 0.25$ are minor mergers and $\mu_\star <0.1$ are mini mergers. 

\begin{table}[!htb]
\begin{tabular}{lccc}
\hline\hline
Simulation & $\rm N_{\rm hosts}$ & $H(z)$ & $H(M_\star)$ \\
\hline
\tng      & 477 & 0.061 & 0.051  \\
\lenient  & 926 & 0.000 & 0.036 \\
\strict   & 485 & 0.000 & 0.026 \\
\hline
\end{tabular}
\caption{The number of MBH pair host samples in the simulations. Each MBH pair host is matched to a unique non-merging control galaxy in redshift and stellar mass, so the number of controls equals $N_{\rm hosts}$ in every run. $H(z)$ and $H(M_\star)$ are the Hellinger distances (Eq.~\ref{eq:Hellinger}) between the pair-host and control distributions in redshift and stellar mass.}
\label{table:sim-details}
\end{table}

\section{Results}
\label{sec:results}

\subsection{Comparison of MBH pair hosts with the full \tng\ and \brahma\ galaxy populations}
\label{sec:MBHpairstofullpopulation}
\begin{figure*}
    \centering
    \includegraphics[width=0.9\linewidth]{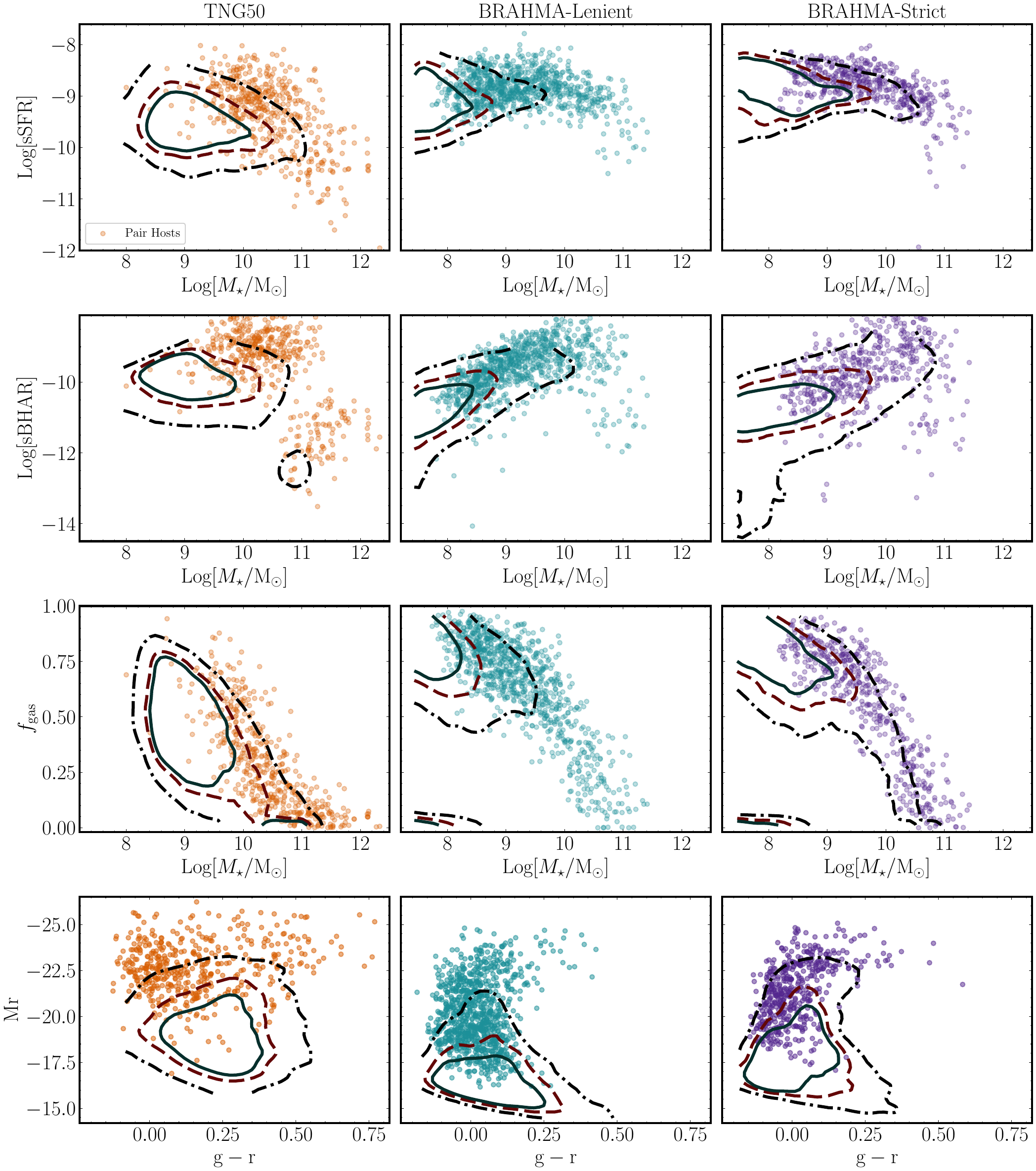}
    \caption{Properties of the pair hosts (dots) compared to the full subhalo population across all snapshots. Contours represent 50\% (solid), 68\% (dashed), and 90\% (dot-dashed) percentiles of the subhalo population for \tng\ (\textit{left column}), \lenient\ (\textit{middle column}) and \strict\ (\textit{right column}). \textit{Rows from top to bottom:} specific star formation rate (sSFR), specific black hole accretion rate (sBHAR), gas fraction ($f_{\rm gas} = M_{\rm gas}/(M_{\rm gas}+M_*)$), and rest-frame color--magnitude diagram ($M_r$ vs.\ $g-r$). The pair hosts occupy the more massive and brighter part of the galaxy population, also highlighting the need for mass-matched controls when comparing their properties to non-mergers.}
    \label{fig:mergers_vs_subhalos}
\end{figure*}

Before quantifying merger-driven enhancements, we first establish the baseline properties of MBH pair hosts relative to the full galaxy population and highlight relevant differences between the \tng\ and \brahma\ simulations. Figure \ref{fig:mergers_vs_subhalos} shows the distribution of MBH pair hosts (colored points) overlaid on the full subhalo population, comprising both pair hosts and non-mergers over all snapshots (contours), for the three simulations: \tng, \lenient\ and \strict. We compare the specific black hole accretion rate (sBHAR), the specific star formation rate (sSFR), the gas fraction (\fgas = $M_{\rm gas}/(M_{\rm gas}+M_{\rm \star})$), and the color--magnitude diagram of the MBH pair hosts to those of the overall population. In all three runs MBH pair hosts occupy more massive subhalos than the full population, with the median \Mstar\ of the pair hosts lying 1.19, 1.21 and 1.28 dex above the median subhalo \Mstar\ in \tng, \lenient\ and \strict\ respectively. This mass offset is what drives much of the apparent separation in sSFR, sBHAR and \fgas. 

The sSFR of pair hosts declines with stellar mass in \tng\ where the pair-host median drops by 1.5 dex between $10^{9}$ and $10^{12}$ \msun. In the \brahma\ runs the decline in sSFR with stellar mass is milder and only starts dropping above $10^{10}$ \msun\ with \lenient\ showing a drop of 0.6 dex and \strict\ showing a drop of 0.8 dex across the stellar mass range they cover. The sBHAR shows bimodality in \tng, where below a stellar mass of $M_\star \sim 10^{11}$ \msun\ the pair hosts show $\mathrm{log[sBHAR]} \simeq -9.4$ and above which the pair hosts fall to a lower value of $\mathrm{log[sBHAR]} \simeq -11.5$. This split reflects the two BH feedback modes in \tng, with the lower-sBHAR population arising from the kinetic feedback mode that sets in at high BH masses. In both the \brahma\ runs sBHAR rises with stellar mass but declines at the highest $M_\star$ values. The \fgas\ also decreases for higher stellar mass hosts, but the gas fractions in \lenient\ and \strict\ pair hosts are generally higher than in \tng.

The color--magnitude panel in Figure \ref{fig:mergers_vs_subhalos} shows that across the three simulations, MBH pair hosts are brighter than the overall population by $3.3-3.5$ magnitudes in $M_r$, consistent with their higher masses. In \tng, the median pair host is slightly bluer than the overall subhalo population by $0.12$ in $g-r$ compared with only $\sim 0.05$ in \brahma. At fixed redshifts, the median $g-r$ offset in \tng\ is $\lesssim 0.04$ mag and below $z=0.5$ the pair hosts are marginally redder than the general population. A tail of redder, brighter pair hosts is thus visible in \tng, plausibly corresponding to quenched massive MBH pair hosts \citep{cella_host_2025}. $6 \%$  of \tng\ pair hosts have $g-r >0.5$, and all of them are below $z=1.5$ with a median $z=0.34$. This red branch comprises massive galaxies with a median  $\log_{10} (M_{\star}/M_{\odot}) = 11.5$, and its median sSFR lies $1.8$ dex below that of the other pair hosts. This branch is essentially absent in \brahma\ which is expected as those runs only reach $z=1.3$.

\begin{figure}[!htb]
    \centering
    \includegraphics[width=0.95\linewidth]{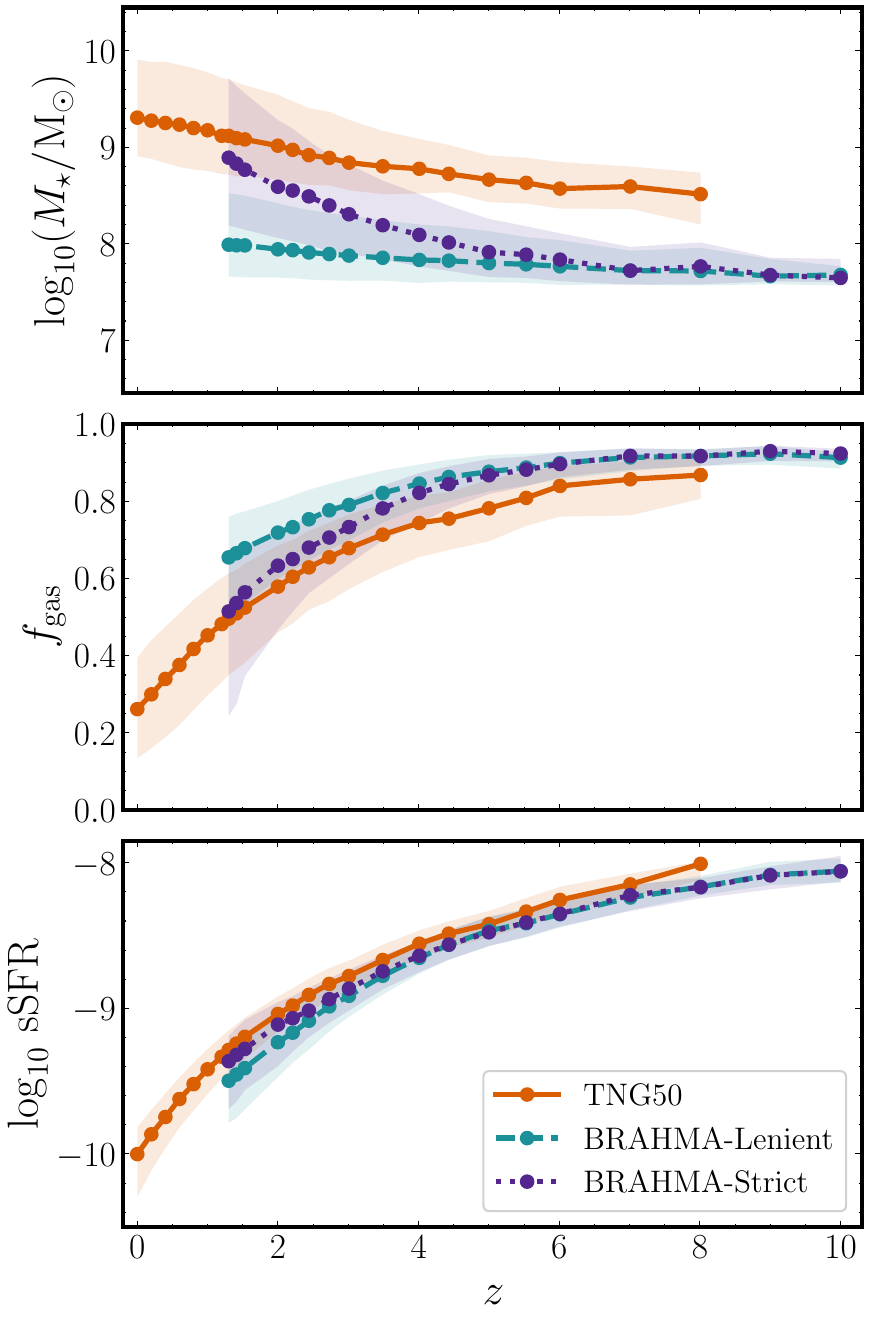}
    \caption{The comparison of properties (\fgas, \Mstar, and sSFR) across the redshift range of their respective simulations for \tng, \lenient\ and \strict\ for subhalos selected in this study. The \brahma\ simulations have higher median \fgas\ compared to \tng\ with agreement within their uncertainties closer to lower redshifts. The stellar masses probed by the simulations are a result of the selection choices while sSFR shows good agreement across the three simulations. }
    \label{fig:sim-subhalo-comparison}
\end{figure}

We have further compared the three simulations in Figure \ref{fig:sim-subhalo-comparison} which shows the redshift evolution of stellar masses, \fgas\ and sSFR. The median stellar mass differs by about a dex between \tng\ and \lenient, and this arises mainly due to the different selection criteria between the two simulations as detailed in Section \ref{sec:subhalo-selection}. This includes the different stellar-particle number cut adopted for \tng\ and \brahma, and the requirement that each selected subhalo host at least one MBH. This condition raises the median stellar mass in \tng\ because of its higher BH seeding threshold compared to \lenient. \strict\ lies between the two, as expected from seeding criteria that are more restrictive than \lenient. In all runs, \fgas\ increases with redshift and exceeds 0.8 at $z\gtrsim 6$, though \fgas\ is $0.02-0.16$ higher in \brahma\ than in \tng\ across the redshift range. The sSFR distributions in the three simulations show good agreement, with the median sSFR differing by only 0.2 dex, with \tng\ being marginally higher, consistent with all three runs having the same star formation physics. 

\subsection{Enhanced sSFR and sBHAR in MBH pair hosts}
\label{sec:sSFR-sBHAR-enhancements}

We now compare each pair host with its redshift- and stellar-mass-matched non-merging control galaxy. The excess specific star formation rate (sSFR) of MBH pair hosts relative to non-merger controls is defined as

\begin{align}
\Delta \log (\mathrm{sSFR}) &=
\log_{10}\!\left(\mathrm{sSFR}_{\mathrm{merger}}\right)
\nonumber \\
&\quad - \log_{10}\!\left(\mathrm{sSFR}_{\mathrm{control}}\right).
\end{align}

\noindent Similarly for sBHAR,

\begin{align}
    \Delta \log (\mathrm{sBHAR}) &=   
\log_{10}\!\left(\mathrm{sBHAR}_{\mathrm{merger}}\right)
\nonumber \\
&\quad - \log_{10}\!\left(\mathrm{sBHAR}_{\mathrm{control}}\right).
\end{align}

\noindent A positive value indicates enhancement in the pair host relative to its control. 

\begin{figure}[!htb]
    \centering
    \includegraphics[width=0.95\linewidth]{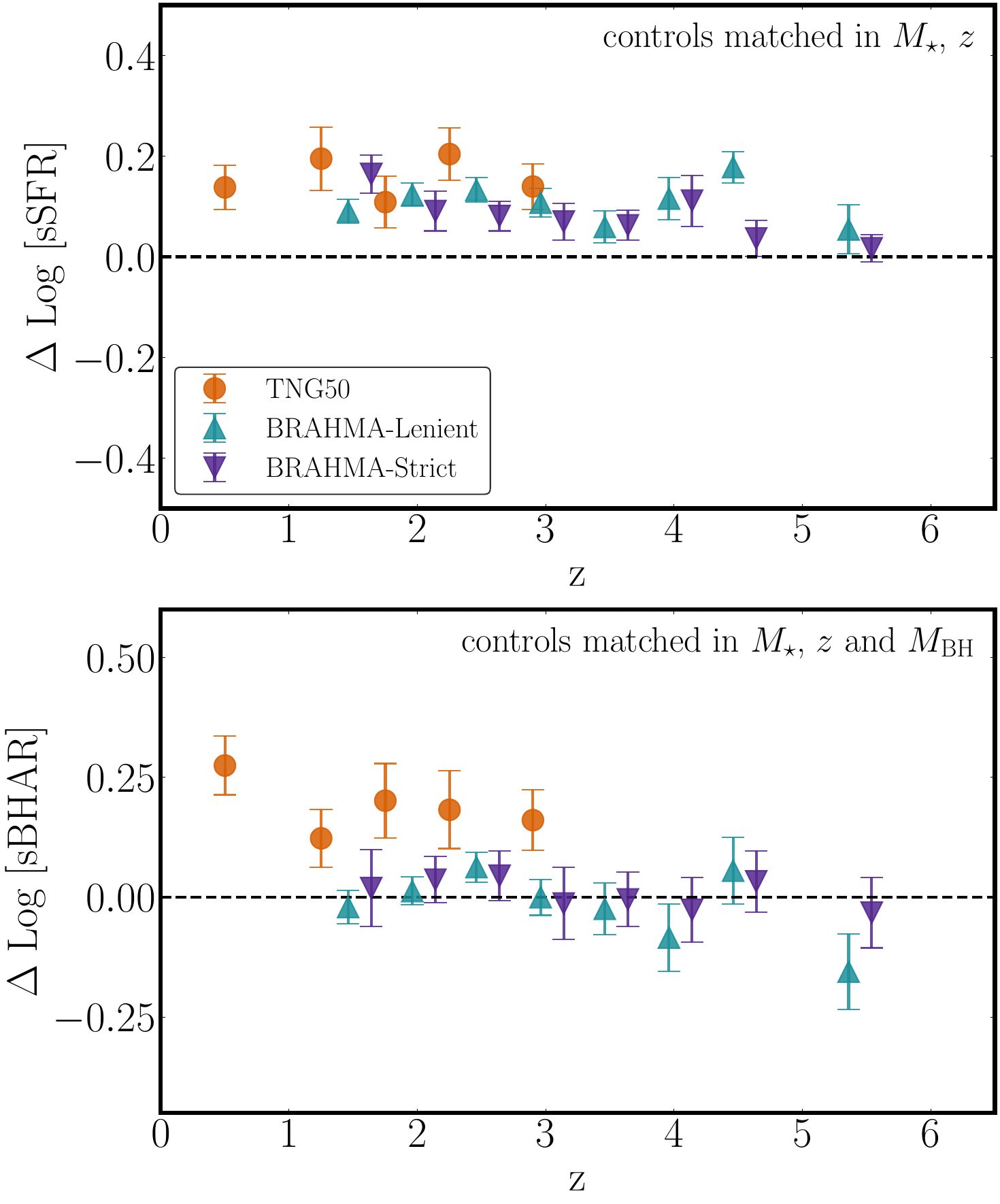}
    \caption{sSFR (top) and sBHAR (bottom) enhancements across redshift for \tng\ and \brahma\ for MBH pair hosts. Each point is the median of the host--control enhancement with its standard error and the dashed line indicates no enhancement. Controls are matched in \Mstar\ and $z$ for the sSFR enhancement, with additional matching in $M_{\rm BH}$ to within $0.1$~dex applied for the sBHAR enhancement measurement. The redshift bins are adaptive, widening until they hold a minimum of 50 pair hosts for \tng\ and 20 pair hosts for \brahma. Pair hosts show positive median sSFR offsets in all three simulations, whereas a clear positive sBHAR offset is detected only in \tng.}
    \label{fig:sSFR-and-sBHAR-enhancement}
\end{figure}

The top panel of Figure \ref{fig:sSFR-and-sBHAR-enhancement} shows the median sSFR enhancement and standard error across the redshift ranges covered by the three simulations. In \tng, over $0\leq z < 3.3$, we measure $\Delta\log({\rm sSFR}) = 0.14 \pm 0.02$~dex, which remains flat across the redshift range. The \brahma\ runs extend down to $z=1.3$, so the three simulations overlap in  $1.3\leq z < 3.3$ where they show good agreement in sSFR enhancement. The \lenient\ run gives an overall enhancement of $0.11 \pm 0.01$ dex and \strict\ gives $0.09 \pm 0.01$ dex over $1.3\leq z < 6$. In \lenient\ the sSFR enhancement is flat with redshift, between $0.09 - 0.12$~dex, whereas in \strict\ it declines from $0.17 \pm 0.04$~dex at $z = 1.3$--$2$ to $0.05 \pm 0.02$~dex at $z = 4$--$6$. The sSFR enhancement therefore declines slightly toward the highest-redshift bins but remains slightly positive throughout. 

The bottom panel of Figure~\ref{fig:sSFR-and-sBHAR-enhancement} shows the analogous redshift trend in sBHAR enhancement. As sBHAR is normalized by BH mass, we keep only those pair hosts with controls matched in BH mass  within $0.1$ dex, in addition to the matching done in stellar mass and redshift. In \tng\ we find $\Delta\log({\rm sBHAR})$ to slightly decline from $0.22 \pm 0.04$ dex in $z=0-2$ to $0.17 \pm 0.05$ dex in $z=2-3.3$. We measure no significant sBHAR enhancement in the \brahma\ runs with $0.005 \pm 0.015$~dex in \lenient\ and $0.008 \pm 0.023$~dex in \strict. 

An important caveat in comparing the sBHAR enhancement between \tng\ and \brahma\ is that the two simulations probe different BH mass ranges and treat BH seeding and dynamics differently. \tng\ uses repositioning, which pins the BH to the potential minimum, whereas the \brahma\ runs track BHs with the subgrid dynamical friction model and retain multiple BHs per subhalo. In this case, sBHAR is measured for the most massive BH in the subhalo, as detailed in  Section \ref{sec:methods}. In \brahma, this BH is in many cases close to the seed mass and has grown very little. Over the redshift range where these simulations overlap ($1.3 \leq z < 3.3$), $\sim 6\%$ of pair hosts in \lenient\ and $\sim 19\%$ in \strict\ have a most massive BH still within $10\%$ of the seed mass $M_{\rm seed}$, and the median BH masses in pair hosts are $\log(M_{\rm BH}/M_{\odot}) =6.2$ and $5.7$ respectively. The BH repositioning scheme in \tng\ may also raise post-merger accretion relative to the subgrid-DF prescription in \brahma, as gas densities will generally be highest in galactic centers. We have tested this directly in Appendix \ref{Appendix1} using a pair of \brahma\ runs that share the \lenient\ seeding criteria and mass resolution but differ in the BH dynamics. We find that the run that uses BH repositioning raises $\Delta\log({\rm sBHAR})$ only by $\lesssim 0.08$ dex.

\subsection{AGN excess in MBH pair hosts}
\label{sec:AGNexcess}

The sBHAR enhancements, or lack thereof, motivate the question of how merger-driven accretion translates into AGN luminosity. The bolometric AGN luminosity of a BH is defined as  

\begin{equation}
    L_{\rm bol} = \frac{\epsilon_r}{1-\epsilon_r} \dot{M}_{\rm BH} c^2
\end{equation}

\noindent where $\epsilon_r$ is the radiative efficiency ($\epsilon_r$ = 0.2 in \tng\ and \brahma) and $\dot{M}_{\rm BH}$ is the BH accretion rate. We first compare the AGN bolometric luminosities of pair hosts to those of their controls. Similar to sBHAR enhancement measurement, we require the controls to be also matched to the pair hosts in $M_{\rm BH}$ with a tolerance of $0.1$ dex. 

In \tng\ the median AGN luminosity offset between pair hosts and their matched controls, for pairs with $L_{\rm bol}>10^{42} \rm \, erg \, s^{-1} $, is $0.22$ dex, so pair hosts are typically $\sim 1.6 \times$ more luminous at $z<3.3$. In \brahma, the same offset is $\sim 0.03$ dex in \lenient\ and $\sim 0.05$ dex in \strict. Thus, pair hosts in \brahma\ have little or no enhancement in bolometric luminosity.  Note that the \brahma\ BHs are less massive than those in \tng, with median BH masses of $\log (M_{\rm BH}/M_\odot) = 6.2$ and $5.7$ compared to $7.6$ in \tng, and are therefore less luminous overall. 

\begin{figure}[!htb]
    \centering
    \includegraphics[width=0.9\linewidth]{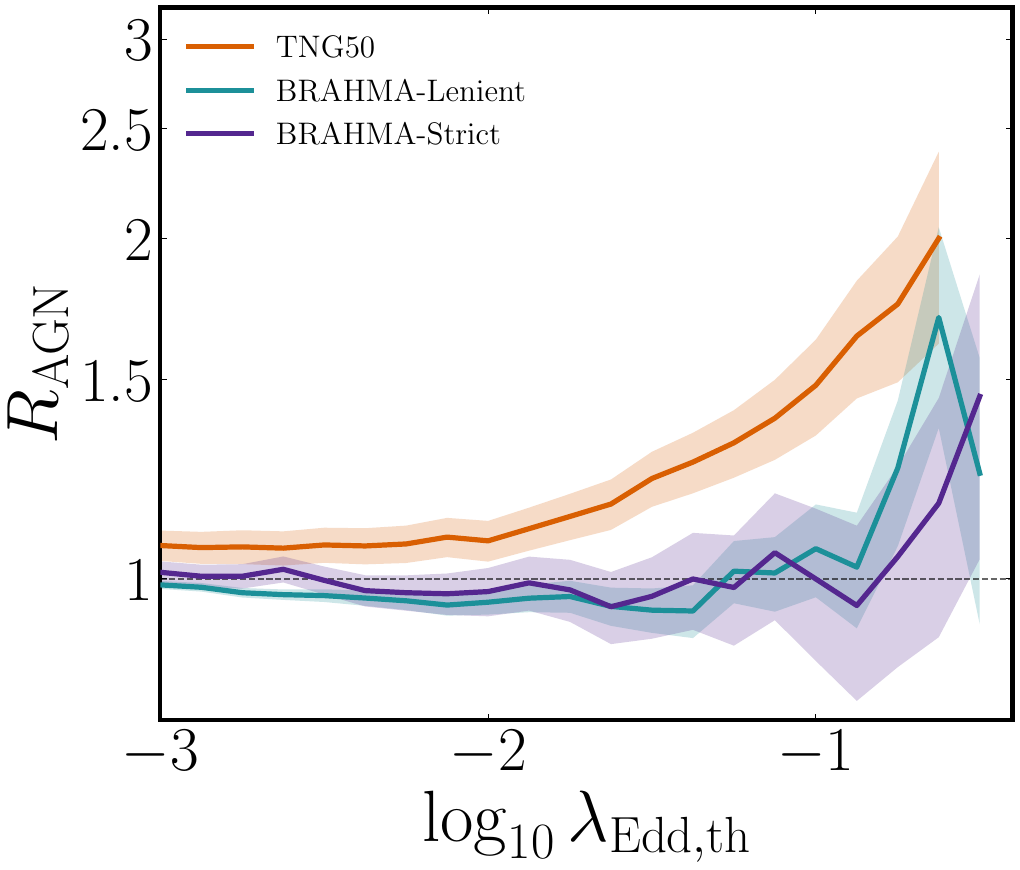}
    \caption{AGN excess in MBH pair hosts relative to controls, defined as the ratio of $f_{\rm AGN}$ (the fraction of sources with $\lambda_{\rm Edd} > \lambda_{\rm Edd,th}$) in mergers to that in controls, for \tng, \lenient\ and \strict. The shaded bands represent $1\sigma$  uncertainties and the dashed line marks no excess. Controls are matched in $M_{\star}$, $z$ and $M_{\rm BH}$ and the excess is plotted only where at least 20 sources lie above the threshold in both the pair host and control samples. The AGN excess is largest for the most rapidly accreting MBH pair hosts.}
    \label{fig:fAGN_excess}
\end{figure}
    
For a mass-independent comparison, we use the Eddington ratio ($\lambda_{\rm Edd}$) to define the AGN fraction $f_{\rm AGN}$ as the fraction of sources with $\lambda_{\rm Edd}$ above a threshold $\lambda_{\rm Edd,th}$. Then, AGN excess is defined as

\begin{equation}
    R_{\rm AGN} (\lambda_{\rm Edd} > \lambda_{\rm Edd,th} ) = \frac{f^{\rm merger}_{\rm AGN}}{f^{\rm control}_{\rm AGN}}.
\end{equation}

\noindent Because $\lambda_{\rm Edd}$ is undefined without a BH, the AGN excess is computed only for those pairs where the control non-merger also hosts an accreting BH.

Figure~\ref{fig:fAGN_excess} shows $R_{\rm AGN}$ as a function of $\lambda_{\rm Edd,th}$ for all three runs. In \tng, $R_{\rm AGN}$ rises monotonically with threshold, from $1.08 \pm 0.04$ at $\lambda_{\rm Edd,th} = 0.01$ to $2.00 \pm 0.39$ at $\lambda_{\rm Edd,th} = 0.24$. Thus the relative occurrence of AGN in \tng\ pair hosts compared with controls increases with Eddington-ratio threshold. This is consistent with observations finding the most luminous AGN preferentially in disturbed systems \citep{Triester2012,Fan2016} and simulations reporting the same \citep{McAlpine2020,Byrne_Mamahit_2023}. Neither \brahma\ run shows any significant excess over most of the $\lambda_{\rm Edd,th}$ range. In both runs, $R_{\rm AGN}$ remains between $0.94-1.06$ below $\lambda_{\rm Edd,th} \simeq 0.1$ and rises only weakly above this threshold, with \lenient\ reaching $1.70 \pm 0.34$ at $\lambda_{\rm Edd,th} = 0.24$ and \strict\ reaching $1.45 \pm 0.41$ at $\lambda_{\rm Edd,th} = 0.32$. AGN excess in pair hosts is therefore largely absent in \brahma, with at most a weak signal among the most rapidly accreting MBHs.

\subsection{Dependence on gas fraction, stellar mass and BH mass}
\label{sec:gasfraction-stellarmass-bhmass} 

Having investigated the enhancements in sSFR, sBHAR and AGN luminosity in MBH pair hosts, we now examine how these enhancements vary with gas fraction, host stellar mass and BH mass. To characterize the gas reservoir entering the merger, we define the combined progenitor gas fraction (\fgasprog) following \citet{Hani2020} and \citet{Schechter2025}:

\begin{equation}
    \label{eq:fgasprog}
    f_{\rm gas, prog} = \frac{\sum_{\rm prog} M_{\rm gas}}{\sum_{\rm prog} M_{\rm gas} + \sum_{\rm prog} M_{\star}}
\end{equation}

\noindent Here, the sums run over both progenitor subhalos, with gas and stellar masses measured within twice the stellar half-mass radius of each progenitor. Note that this quantity differs from the descendant host gas fraction introduced in Section \ref{sec:MBHpairstofullpopulation}, which we denote by \fgaspm\ here. 

We distinguish between two gas-fraction offsets to test the gas-richness of pair hosts relative to controls:

\begin{align}
    \Delta f_{\rm gas,prog}
    &\equiv \left\langle f_{\rm gas,prog}
    - f_{\rm gas,control} \right\rangle, \\
    \Delta f_{\rm gas,post}
    &\equiv \left\langle f_{\rm gas,post}
    - f_{\rm gas,control} \right\rangle.
\end{align}

The brackets here represent the mean over matched host-control pairs within each redshift bin. Both offsets use the same controls and their instantaneous gas fractions. The progenitor offset, $\Delta f_{\rm gas,prog}$, characterizes the gas richness of the combined merger reservoir relative to the controls. The descendant offset, $\Delta f_{\rm gas,post}$, directly tests whether the pair hosts are more gas-rich than their controls at the matched redshifts.

\begin{figure}[!htb]
    \centering
    \includegraphics[width=0.95\linewidth]{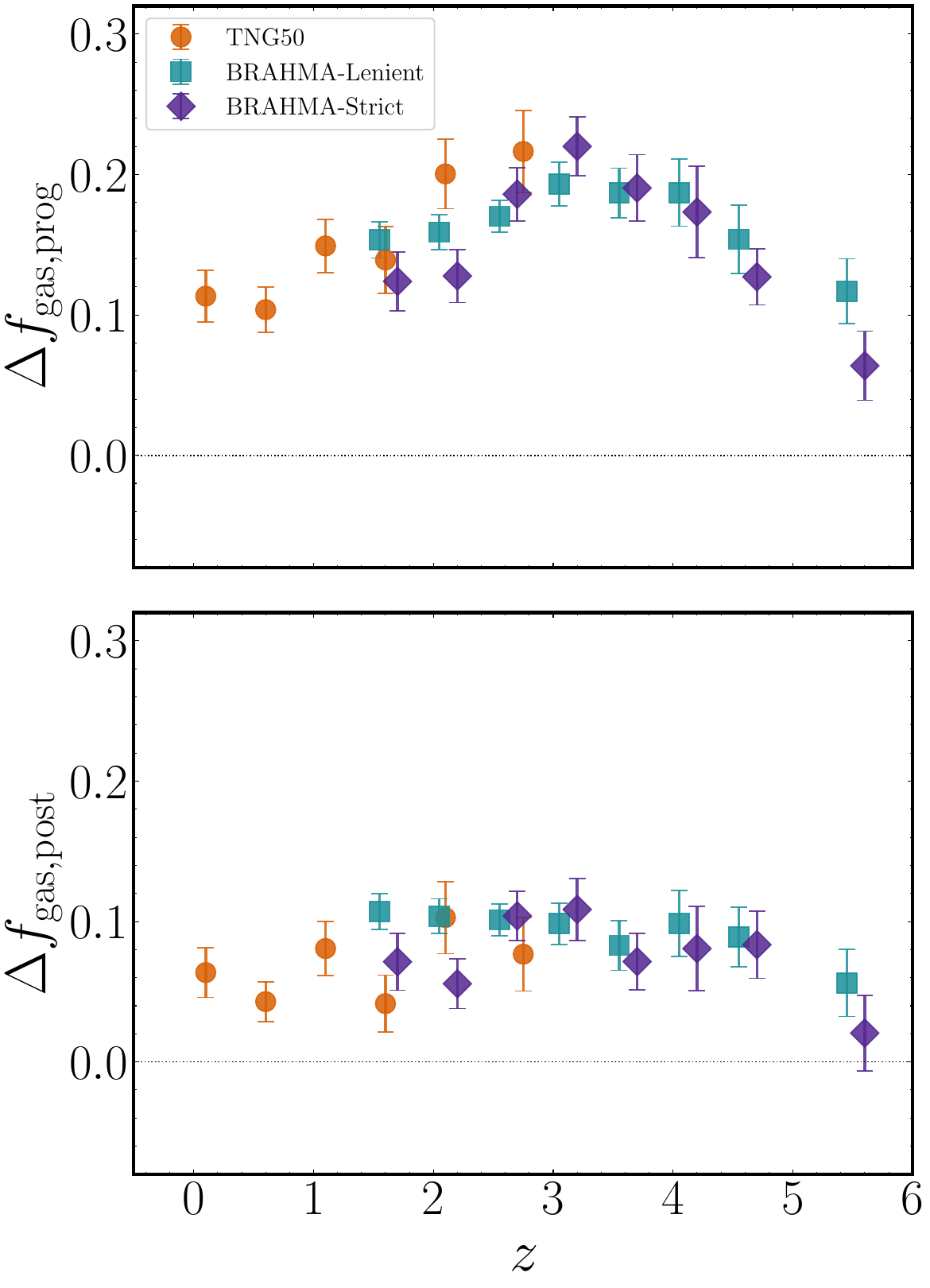}
    \caption{Mean per-pair gas fraction offsets relative to matched controls as a function of redshift for \tng\ and \brahma. \emph{Top:} combined progenitor offsets, $(\Delta f_{\rm gas,prog})$. \emph{Bottom:} descendant host offsets, $(\Delta f_{\rm gas,post})$. Both panels use the same matched host-control pairs, and error bars show the standard error of the gas fraction offsets. The redshift bins contain a minimum of 50 pair hosts for \tng\ and 25 in each \brahma\ run. The three runs overlap in the redshift range $1.3 \leq z < 3.3$. Descendant pair hosts have smaller gas-fraction offsets than their combined progenitors, and the offset weakens in the highest redshift bins.}
    \label{fig:fgas_merger_vs_control}
\end{figure}

Figure~\ref{fig:fgas_merger_vs_control} compares the progenitor-control gas-fraction offsets, $\Delta f_{\rm gas,prog}$ in the top panel and the descendant-control offsets, $\Delta f_{\rm gas,post}$ in the bottom panel for \tng\ and \brahma. In \tng, $\Delta f_{\rm gas,prog}$ is positive in every redshift bin and generally increases with $z$, from $0.11$ at $z \simeq 0$ to $0.22$ at $z \simeq 3$. Since the controls themselves are gas poor, with mean $f_{\rm gas,control}$ below $0.3$, these offsets are substantial relative to their gas content. \lenient\ and \strict\ runs on the other hand show a non-monotonic behavior, with $\Delta f_{\rm gas,prog}$ rising from  $0.12-0.15$ at $z \simeq 1.5$ to a peak of $0.19-0.22$ at $z \simeq 3$ before falling to $0.06-0.12$ by $z \simeq 5$. The decline at the highest redshifts is consistent with an already gas-rich population, as the contrast between progenitor and control gas fractions becomes smaller. $\Delta f_{\rm gas,prog}$ shows no significant dependence on the seed model with both the \lenient\ and \strict\ runs agreeing within their uncertainties in every redshift bin. In the redshift range where the three runs overlap, $1.3 \leq z < 3.3$, all three runs agree, with $\Delta f_{\rm gas, prog} = 0.16$--$0.18$.

The descendant offsets are smaller than the progenitor offsets which may partly reflect the gas evolution between the progenitor and descendant measurement. In \tng, $\Delta f_{\rm gas,post}$ ranges from $\approx 0.04-0.10$, with a weak overall increase towards higher redshifts. In \lenient\ the descendant offset remains $\approx 0.08-0.11$ below $z \simeq 4.5$, while \strict\ shows a less regular trend over $\approx 0.06-0.11$. In the highest redshift bin, the offsets decrease to $0.06\pm0.02$ for \lenient\ and $0.02\pm0.03$ for \strict. Thus, the gas-fraction excess in pair hosts relative to controls becomes smaller at high $z$. Over $1.3\leq z<3.3$, where the three runs overlap, the descendant offsets are $\Delta f_{\rm gas,post} \approx 0.07-0.1$, so descendant pair-hosts remain more gas-rich than their controls.

\begin{figure}[!htb]
    \centering
    \includegraphics[width=0.95 \linewidth]{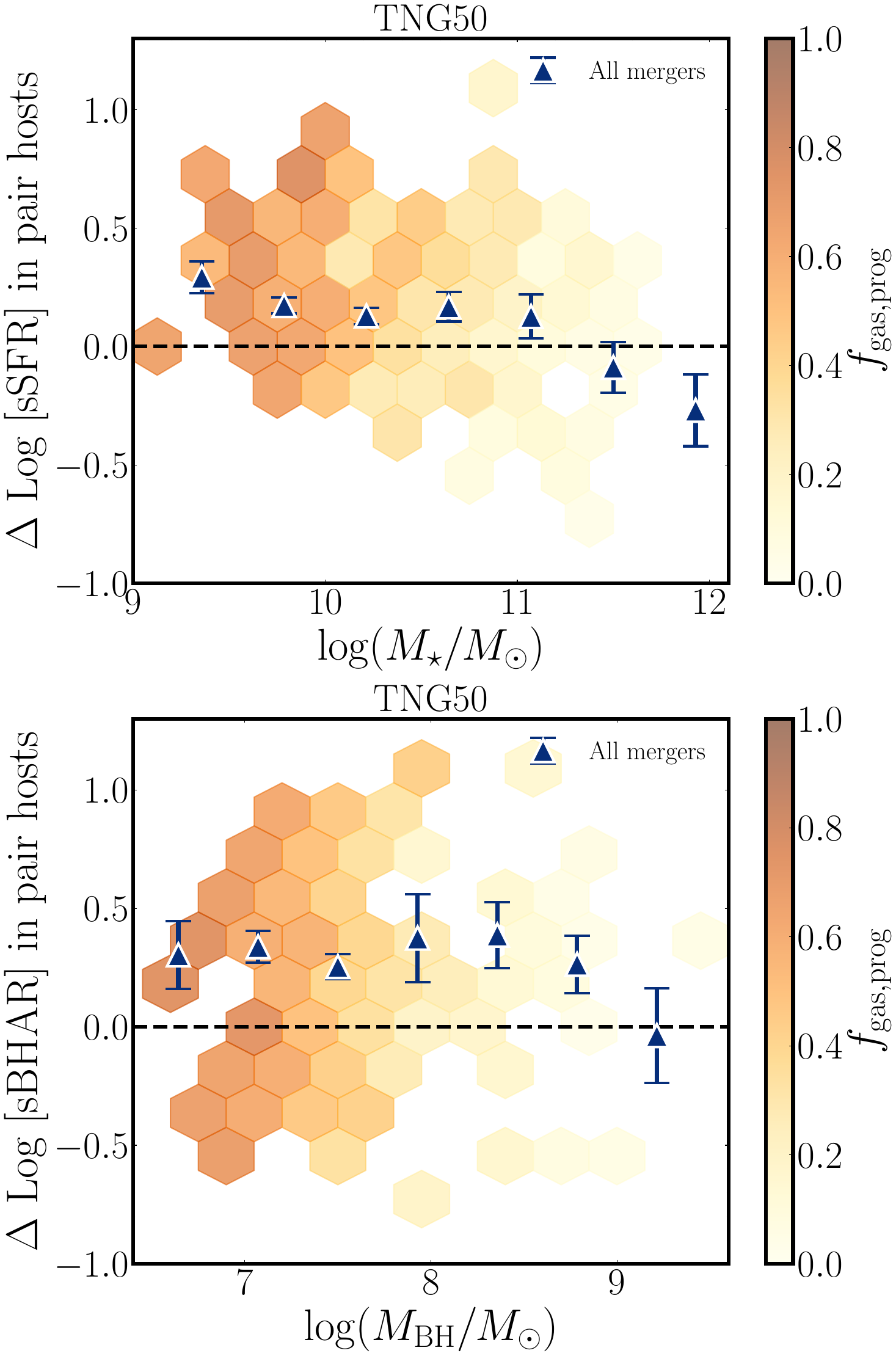}
    \caption{\textit{Top:} sSFR enhancement as a function of merger-host stellar mass in \tng\ where the background hexbin shows the mean $f_{\rm gas,prog}$ of the pair hosts falling in each $\Delta\log({\rm sSFR})$--$M_\star$ cell. \textit{Bottom:} sBHAR enhancement as a function of BH mass in the same sample, with the mean progenitor gas fraction indicated by the hexbins. The triangles represent the median enhancement per mass bin with the standard error and are plotted only where a bin holds at least 10 pairs. The dashed line indicates no enhancement. The sSFR enhancement generally decreases with stellar mass and sBHAR enhancement falls toward the highest BH masses, with declining $f_{\rm gas,prog}$.}
    \label{fig:sSFR-enhancement-w-stellar-mass-sBHAR-enhancement-w-BH-mass}
\end{figure}

The sSFR enhancement also varies with the stellar mass of the pair host \citep{Schechter2025}. The top panel of Figure~\ref{fig:sSFR-enhancement-w-stellar-mass-sBHAR-enhancement-w-BH-mass} shows that the median $\Delta\log({\rm sSFR})$ in \tng\ decreases from $\sim 0.3$~dex at $M_\star \sim 10^{9} M_\odot$ to $\sim 0.1$~dex  at $\sim 10^{11}~M_\odot$, with suppression in the highest stellar mass bin. The background hexbin shows the corresponding mean \fgasprog\ value in the $\Delta \log_{\rm sSFR} - M_{\star}$ plane. The decline in enhancement with \Mstar\ is accompanied by the decline in the progenitor gas fraction, with the mean \fgasprog\ dropping from $0.62$ at $M_\star \sim 10^{9}~M_\odot$ to $0.15$ at $M_\star \sim 10^{11}~M_\odot$. 

The bottom panel of Figure~\ref{fig:sSFR-enhancement-w-stellar-mass-sBHAR-enhancement-w-BH-mass} shows the sBHAR enhancement as a function of BH mass. The enhancement stays flat between $0.25$--$0.40$~dex over BH masses of $\sim 10^{7}M_\odot$--$10^{9}M_\odot$. For the population of MBH pair hosts with $M_{\rm BH} > 10^{9}~M_\odot$, the enhancement drops to $\sim -0.04$~dex, which tracks the decline in the progenitor gas-fraction. In \brahma, we observe the sSFR enhancement stays between $0.05-0.31$ dex across $M_{\star} = 10^{8}$--$10^{11}~M_\odot$ without a monotonic decline, and we find no significant sBHAR enhancement at any BH mass.

\begin{figure*}[!htb]
    \centering
    \includegraphics[width=0.9\linewidth]{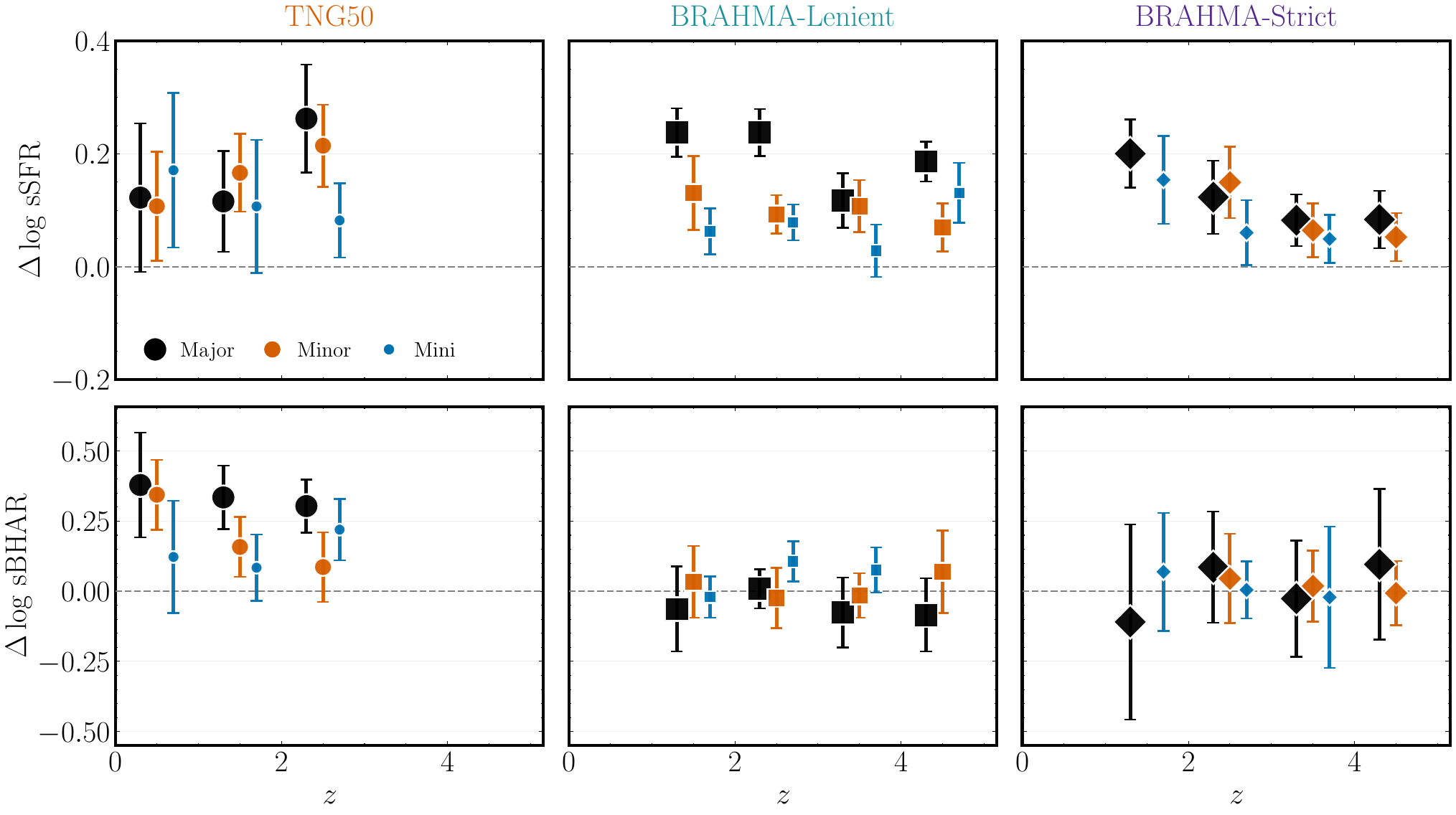}
    \caption{sSFR (top) and sBHAR (bottom) enhancements varying with redshift, split by galaxy merger mass ratio $\mu_{\star}$ into major ($\mu_{\star} \geq 0.25$, black), minor ($0.1 \leq \mu_{\star} < 0.25$, orange) and mini ($\mu_{\star}< 0.1$, blue) mergers for \tng, \lenient\ and \strict. The points represent the median and standard error of enhancements in $\Delta z = 1$, plotted at the bin center and offset horizontally by merger type. We require a minimum of 10 pair hosts per merger type per bin. The controls are matched in \Mstar\ and $z$ for
    the sSFR plots, with additional matching in $M_{\rm BH}$ for the sBHAR measurements.}
    \label{fig:effect_of_mass_ratio}
\end{figure*}

\subsection{Effect of galaxy merger mass ratio}
\label{sec:massratio}

We next examine the effect of the galaxy merger mass ratio on the sSFR and sBHAR enhancements. Figure~\ref{fig:effect_of_mass_ratio} shows the enhancements as a function of redshift, split into major, minor, and mini mergers, with one panel per simulation. We plot the median enhancement in $\Delta z = 1$ bins, requiring at least 10 host-control pairs per mass-ratio class in each bin. \tng\ has too few pair hosts to meet this threshold above $z \simeq 3$, while the \brahma\ runs are followed to $z \simeq 5$.

In \tng\ the three mass ratio classes give consistent sSFR merger enhancements with little dependence on mass ratio except at $z \simeq 2.5$, where the major mergers show a slightly higher median than minor and mini mergers. \lenient\ shows an ordering where the median sSFR enhancement for major mergers lies above both other classes in all four of its redshift bins. \strict\ also shows a similar but weak dependence on merger mass ratio with major mergers showing the highest median enhancement in three of its four redshift bins. The dependence of sSFR enhancement on merger mass ratio is therefore weak across all three simulations.

The mass ratio has a different impact on the sBHAR enhancement than on sSFR enhancement. As in the previous sections, the controls are additionally matched in $M_{\rm BH}$. In \tng\ major mergers give the largest median enhancement in every redshift bin, although the three mass ratio categories overlap within their error bars in all but the $1<z<2$ bin, where major and mini mergers are only marginally separated. Neither \brahma\ run produces any significant sBHAR enhancement, with every redshift bin consistent with zero. The mass ratio dependence of sBHAR enhancement is therefore also weak.

\section{Discussion}
\label{sec:discussion}
\subsection{Comparison to other simulations}
\label{sec:comparisontosims}

We now turn to compare our results with existing literature on galaxy merger-driven enhancements in star formation and BH accretion. Several large-volume and zoom-in cosmological simulations have also been used to study merger-based enhancement in a cosmological context. The works of \citet{Patton2020} using \tngonehundred, \citet{Martin2017} using Horizon-AGN, and \citet{Schechter2025} using \tng\ have all shown that sSFR strength in mergers decreases with increasing redshift. \citet{Hani2020} analyzed the post-merger galaxies from \tngthreehundred\ to show that the SFRs are enhanced by a factor of $\sim 2$ compared to controls in $ 0 \leq z \leq 1$ with no evident redshift evolution. Other works also show an SFR enhancement factor of $\sim 2-3$ in this low redshift range \citep{Montero2019,Patton2020}. Our sSFR enhancement results qualitatively agree with all the results quoted here, with pair hosts (equivalent in our analysis to post-merger galaxies) effectively having $\sim 1.4 \times$ sSFR in mergers for $0\leq z \leq 3.3$ (see Figure \ref{fig:sSFR-and-sBHAR-enhancement}). 

Our work extends the sSFR enhancement measurement to higher redshifts and more detailed BH seeding models with the \brahma\ runs. We show that the sSFR--merger connection survives at higher redshifts ($z \gtrsim 3$), although it is weaker. We also observe that the sSFR enhancement decreases with increasing host stellar mass in Figure \ref{fig:sSFR-enhancement-w-stellar-mass-sBHAR-enhancement-w-BH-mass} as also seen in other works \citep[e.g.,][]{Schechter2025}. A similar trend is also seen in the BH merger hosts of \citet{degraf_morphological_2021}, where the elevated star formation is found only in low-mass host mergers, with high-mass hosts showing typical star formation rates due to their low gas fractions and strong AGN feedback. 

The picture for BHAR enhancement in cosmological simulations is more mixed. \citet{Byrne_Mamahit_2023,Byrne_Mamahit_2024} used \tngonehundred\ to investigate sBHAR enhancement in post-mergers and found an average enhancement of $0.23$ dex in the redshift range $0<z<1$. They also observed that only $\sim 35 \%$ of post-mergers experience a luminous AGN with $L_{\rm bol} > 10^{45} \rm erg \, s^{-1}$. Their results show that while mergers can trigger AGN, the majority of galaxies hosting high-luminosity AGN are not recent mergers. \citet{Schechter2025} use \tng\ to probe sBHAR enhancements via mergers up to $z \sim 3$ and observe a decrease in the enhancement with increasing redshifts. \citet{Sharma2024} and \citet{Jhee2026} use zoom-in simulations to study the merger--AGN connection to see that mergers are not the primary drivers of AGN activity except for certain cases at low redshifts. They note that this is the regime where galaxies are more gas-poor and where mergers play a more important role in fueling BH accretion. Our results with \tng\ at lower redshifts agree with \citet{Schechter2025} and \citet{Byrne_Mamahit_2023,Byrne_Mamahit_2024}. We observe an sBHAR enhancement of $0.22$ dex at $0<z<2$ with \tng, which slightly declines with redshift. However, we do not observe any significant sBHAR enhancement in \brahma, and this remains true at high redshifts. We also note that the fraction of pair hosts with $L_{\rm bol} > 10^{45} \rm erg \, s^{-1}$ is $33 \%$ in \tng, 9\% in \lenient, and 14\% in \strict\ over $1.3 \leq z < 3.3$. Thus, luminous AGN occur in only a minority of pair hosts in all three runs, indicating a recent galaxy merger between BH-hosting progenitors does not necessarily coincide with high AGN activity.

Simulations have also examined the dependence of sSFR/sBHAR enhancement on the mass ratio of the galaxy merger. While works such as \citet{kaviraj_galaxy_2015} find major mergers increase star formation by $20-40 \%$, others such as \citet{Martin2017} find that both major and minor mergers enhance SFR in comparable amounts at any given redshift. Using \tngthreehundred, \citet{Hani2020} showed that minor mergers ($0.1 \leq \mu \leq 0.3$) contribute to $\sim 50 \%$ of the merger driven enhancements in $0\leq z\leq1$. On the other hand, \citet{Schechter2025} using \tng\ find that minor mergers generally have smaller sSFR enhancements compared to major mergers and also find the highest sBHAR excess in major mergers. \citet{Byrne_Mamahit_2023} showed that the gas-rich major mergers provided a consistent sBHAR enhancement but not necessarily greater enhancement compared to minor mergers. \citet{byrne-mamahit_interacting_2025} looked at mini mergers and  their AGN triggering and found that the AGN excess can exist for post-merger galaxies with merger mass ratios as low as $\mu = 1/40$. Our results only show a weak mass-ratio dependence in both quantities. \lenient\ and \strict\ runs show an ordering in sSFR enhancement, with major mergers having the highest median. \tng\ on the other hand has all three mass ratio classes produce similar levels of sSFR enhancement. Major mergers in \tng\ do show a higher sBHAR median than minor and mini mergers, but the three classes lie within their uncertainties.

\subsection{Comparison to observations}
\label{sec:comparisontoobservations}
On the observational side, numerous SDSS-based studies have established that the galaxy pairs with close companions show  enhanced levels of star formation, at least at low to moderate redshift \citep{Barton2000, Kewley2006, Ellison2008, Woods2010, Scudder2012, Patton2013}, accompanied by enhanced molecular \citep{Violino2018} and atomic \citep{Ellison2018} gas reservoirs. Post-merger galaxies also show a measured central SFR enhancement by a factor of $\sim 2.5-3.5$ compared to isolated control galaxies in the local universe \citep{Ellison2013, Thorp2019}. We measure a lower enhancement of $ \sim 1.4 \times$ over $0\leq z < 3.3$, though our measurement is the total star formation rate of the host rather than the central star formation rate which is where the observed enhancement is concentrated. \citet{Willett2015} showed that major and minor mergers impact star formation at similar levels, which is consistent with our prediction of weak mass-ratio dependence of the sSFR enhancement.

The merger--AGN connection is observationally more contested, with the strength of the inferred connection depending strongly on the selection wavelength and redshift range probed. Studies using galaxy pairs in SDSS have found an excess AGN fraction relative to control galaxies that increases with decreasing pair separations \citep{Ellison2011,Ellison2013}. \citet{Satyapal2014} measured a $\sim 10-20$ excess of mid-IR selected AGN in SDSS post-mergers matched to WISE, while soft X-ray selected samples at $z < 1$ show no excess of merger morphologies relative to inactive controls \citep{Grogin2005, Cisternas2011, Kocevski2012, Villforth2014}. Obscuration is likely a substantial part of this disagreement. \citet{Ellison2019} found $\sim 30\%$ of optically selected AGN hosts to be morphologically disturbed compared to $\sim 60\%$ of mid-IR selected hosts, and \citet{Weston2017} found WISE-selected AGN to be $5\times$ more likely to be obscured when the host is a post-merger galaxy. \citet{Koss2010} selected AGN in the ultra-hard X-ray band with \textit{Swift}/BAT, which is largely unaffected by the attenuation that suppresses soft X-ray signatures, and recovered an excess of merger signatures at $z < 0.05$. \citet{Ricci2017} subsequently showed with the same selection that the obscuring column peaks in late stage mergers. Optical to soft X-ray surveys of merger-triggered AGN are therefore susceptible to incompleteness due to obscuration.

At higher redshift, studies using HST find no evidence that mergers preferentially trigger AGN \citep{Kocevski2012, Mechtley2016, Marian2019}. JWST has extended this comparison into the epoch probed by our \brahma\ runs. Close-pair analyses in JADES detect a weak but significant sSFR excess of $12\pm5\%$ (a factor of $1.12\pm0.05$) at the smallest separations $r_{\rm p}\lesssim20$~kpc, rising toward $z=5$--$9$, but find no significant AGN excess in pairs over $z=3$--$9$ \citep{Puskas2025}. \citet{Duan2026} likewise report merger-enhanced star formation with a weak AGN excess at high redshift, and \citet{Omori2026} find that the interaction-induced starbursts significantly boost stellar mass assembly at $z \sim 5$. Our \lenient\ and \strict\ runs show a similar pattern with non-zero sSFR enhancements at high redshift but sBHAR enhancements consistent with zero. We note, however, that these observations probe pre-coalescence galaxy pairs whereas our sample consists of galaxy-merger descendants selected as candidate MBH-pair hosts. \citet{Jhee2026} show that restricting a simulated sample to pre-merger phases significantly weakens the apparent AGN--merger connection. A like-for-like comparison at high redshifts therefore requires observations targeting post-merger systems, which depends on improved merger identification methods \citep{ferreira2020,guzman-ortega2023}. Overall, our results broadly agree with the current observations that show evidence of AGN enhancement at low redshift but little to none at high redshift. 

\section{Summary and Conclusions}
\label{sec:summary}

In this paper, we investigate the role of mergers in enhancing star formation and BH accretion within galaxies using three cosmological hydrodynamical simulations that primarily differ in how MBHs are seeded and dynamically evolved. First, we consider \tng, the highest-resolution box of the \illtng\ simulation suite, which seeds $1.2\times10^{6}$ \msun\ BHs in massive $\sim7.4 \times10^{10}$ \msun\ halos and employs a repositioning scheme to keep them near the local potential minimum. We then consider two $[36.9 ~\rm Mpc]^3$ boxes from the \brahma\ simulation suite, with resolutions comparable to \tng. In contrast to \illtng, these simulations employ physically motivated seeding models that place $1.5\times10^{5}$ \msun\ seeds in substantially lower-mass halos and at higher redshifts, while their subsequent orbital evolution is followed using a subgrid dynamical friction model. The different simulations offer complementary advantages. \tng\ has a larger volume than the \brahma\ runs, providing better statistical coverage of massive, rarer halos. However, its more restrictive seeding criterion leads to BH formation predominantly at later cosmic times, limiting our analysis to $0 \leq z < 3.3$. In contrast, while the two \brahma\ runs have smaller volumes and have only been evolved down to $z\sim1.3$, their more permissive seeding criteria allow us to probe merger-associated signatures in lower-mass halos and at higher redshifts, spanning $1.3 \leq z < 6$. All three simulations overlap over $1.3 \leq z < 3.3$, where we can directly compare their results. For each pair host in our simulations, we identify a unique non-merging galaxy as its control, matched in redshift and stellar mass, with additional matching in BH mass for the BH-related properties. We compare the general properties of the pair hosts against the full galaxy population in each run and then quantify enhancements relative to the matched controls in specific star formation rate (sSFR), specific black hole accretion rate (sBHAR), and AGN fraction. 

We observe enhanced gas fractions and star formation in MBH pair hosts across all three simulations. The median sSFR enhancement is $\approx 0.09-0.14$ dex in each simulation, and shows little redshift evolution apart from a decline at the highest redshift bins observed for the \brahma\ runs (see Figure \ref{fig:sSFR-and-sBHAR-enhancement}). In both the \brahma\ runs, the sSFR and $f_{\rm gas}$ enhancements persist to the highest redshift probed ($z \sim 6$). In \tng, the sSFR enhancement declines with increasing host stellar mass (see Figure \ref{fig:sSFR-enhancement-w-stellar-mass-sBHAR-enhancement-w-BH-mass}). The sSFR enhancement is also largely insensitive to the merger mass ratio and the merger type as shown in Figure \ref{fig:effect_of_mass_ratio}.

In contrast, the BH accretion response differs sharply between the simulations. In \tng, pair hosts show a clear sBHAR enhancement of $\approx 0.21$ dex with only mild decline toward higher redshift (see Figure \ref{fig:sSFR-and-sBHAR-enhancement}). In the \brahma\ runs however, the sBHAR enhancement is negligible despite the enhancements in $f_{\rm gas}$ and sSFR. In a similar vein, \tng\ hosts show an excess AGN fraction relative to matched controls that increases with Eddington-ratio threshold. In \brahma, this excess is weak and appears only among the most rapidly accreting BHs. We also investigate (in Appendix \ref{Appendix1}) whether the difference in BH dynamics treatments~(repositioning vs subgrid dynamical friction) could explain the difference in the impact of mergers on BH accretion between \tng\ and \brahma. We test this by comparing merger-driven accretion enhancements between two smaller-volume \brahma\ runs that are identical except for BH dynamics: one with repositioning and one with subgrid DF. Over the full sample, the sBHAR enhancement in the repositioning run only differs by at most 0.08 dex from the subgrid DF run, although the effect of repositioning may be larger at higher BH masses.

The difference in sBHAR enhancement between \tng\ and \brahma\ therefore likely reflects differences in the MBH and host-galaxy populations produced by their respective seeding models. The \brahma\ seeding models produce substantially lower-mass MBHs in lower-mass halos. These differences could affect the BH accretion response to mergers through multiple pathways. First, the steep $M_{\rm BH}^{2}$ dependence of the Bondi--Hoyle accretion rate results in systematically slower growth for the \brahma\ seeds, which are $\sim8$ times less massive than those in \tng, and may alter their response to merger-driven changes in the surrounding gas. Second, the lower halo masses of \brahma\ pair hosts imply shallower gravitational potential wells, potentially changing how stellar and AGN feedback regulate the nuclear gas supply during mergers. Finally, the different simulation volumes result in largely distinct host-halo mass distributions, making it difficult to do an even-handed comparison of distinct BH populations in otherwise similar environments. We will investigate these possibilities in future \brahma\ simulations with larger volumes, which will provide greater overlap with the BH and halo masses sampled by \tng, as well as in simulations employing variations of the accretion and feedback models. 

\begin{acknowledgments}
We thank Aimee Schechter, Paul Torrey, and Rainer Weinberger for helpful discussions during the preparation of the work. AKB acknowledges support from NSF-AST 2510738, NSF-AST 2346977 and the NSF-Simons AI Institute for Cosmic Origins which is supported by the National Science Foundation under Cooperative Agreement 2421782 and the Simons Foundation award MPS-AI-00010515. PS thanks the LSST-DA Data Science Fellowship Program, which is funded by LSST-DA, the Brinson Foundation, the WoodNext Foundation, the Research Corporation for Science Advancement Foundation, and Kevin Wells; his participation in the program has benefited this work. PS and LB acknowledge support from NSF awards AST-2307171 and AST-2509457. Anthropic's Claude was used for code review, minor editing and proofreading of the manuscript. The authors reviewed all suggestions and take full responsibility for the final manuscript.
\end{acknowledgments}

\appendix
% \twocolumngrid

\section{Impact of subgrid BH dynamics}
\label{Appendix1}

\begin{figure*}[!htb]
    \centering
    \includegraphics[width=0.98\linewidth]{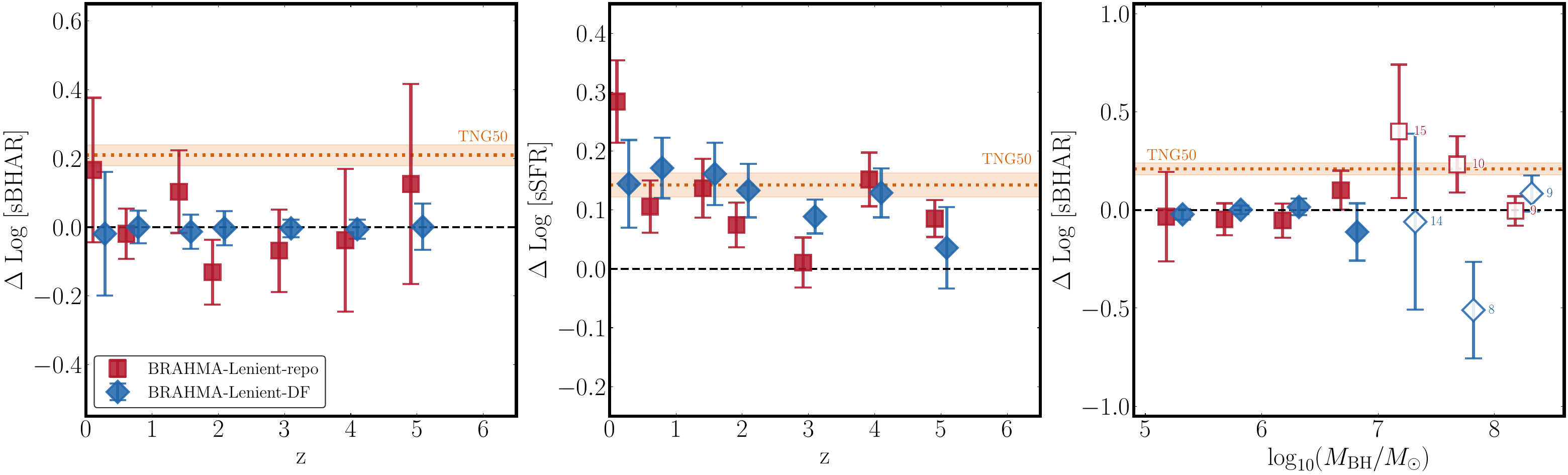}
    \caption{\emph{Left:} the sBHAR enhancement against redshift; \emph{middle:} the sSFR enhancement against redshift; \emph{right:} the sBHAR enhancement against BH mass, for the \lenientrepo\ and \lenientdf\ runs. The dotted orange line is the median enhancement measured in \tng\ over its full redshift and BH-mass range, and the shaded band is its standard error. The two redshift panels use adaptive bins, widened until both runs hold at least $20$ merger-control pairs and the right-hand panel uses fixed $0.5$ dex bins in $\log M_{\rm BH}$, where open markers denote bins holding fewer than $20$ pairs, with the pair count displayed. For the full sample the sBHAR enhancement in \lenientrepo\ exceeds that in \lenientdf\ by at most 0.08 dex, while sSFR enhancement is less affected. For BHs with masses above $10^{7} \rm M_{\odot}$, the sBHAR enhancement in \lenientrepo\ can reach the values observed in \tng, albeit with large uncertainties.}
    \label{fig:bhrepovsdf}
\end{figure*}

To investigate whether BH dynamics affects these results we compare two \brahma\ runs that use the same seeding criteria and the same mass resolution as \lenient\ but in a smaller box of $[ 18.5 \, \rm Mpc]^3$ volume. \lenientrepo\ uses BH repositioning to pin each BH to the nearest potential minimum, while \lenientdf\ evolves the BHs using subgrid dynamical friction. Although the repositioning run's median sBHAR enhancement is higher in several bins, its error bars are several times larger than those of \lenientdf. Over the full sample, \lenientrepo\ raises the median $\Delta\log({\rm sBHAR})$ by at most $0.08$ dex compared to \lenientdf. The sSFR enhancement is less sensitive to the BH dynamics and the median sSFR enhancement differs between the runs only by $-0.04 \pm 0.02 $ dex. For lower mass BHs, below $10^7 \rm M_{\odot}$, both the \lenientrepo\ and \lenientdf\ runs show consistently no significant enhancement. For BH masses above $10^7 \rm M_{\odot}$, where most of the \tng\ pair hosts lie, the \brahma\ samples are sparse. However in this mass range, \lenientrepo\ gives $+0.17 \pm 0.10$ dex against $0.00 \pm 0.19$ dex for \lenientdf, and is closer to the \tng\ sBHAR enhancement value of $0.21 \pm 0.03$ dex. This suggests that repositioning may be responsible for the elevated sBHAR at the highest BH masses, but the sample above $10^{7} M_{\odot}$ is too small for this to be conclusive.

\bibliographystyle{aasjournal}
\bibliography{main}

\end{document}